\documentclass{jfm}
\usepackage{graphicx}
\usepackage{epstopdf, epsfig}
\usepackage{color}
\newcommand{\ct}[1]{\multicolumn{1}{c}{#1}}
\usepackage[normalem]{ulem}
\usepackage{amsmath}

\shorttitle{Interface-resolved simulations of small inertial particles in turbulent channel flow}
\shortauthor{P. Costa, L. Brandt and F. Picano}

\title{Interface-resolved simulations of small inertial particles in turbulent channel flow}

\author{Pedro Costa\aff{1}\corresp{\email{p.simoes.costa@gmail.com}}\thanks{Present address: Faculty of Industrial Engineering, Mechanical Engineering and Computer Science, University of Iceland, Hjardarhagi 2-6, 107 Reykjavik, Iceland}, Luca Brandt\aff{1} \and Francesco Picano\aff{2}
}

\affiliation{
\aff{1}Linn\'e FLOW Centre and SeRC (Swedish e-Science Research Centre), KTH  Mechanics, SE-100 44 Stockholm, Sweden\\
\aff{2}Department of Industrial Engineering, University of Padova, Via Venezia 1, 35131 Padova, Italy
}

\begin{document}

\maketitle

\begin{abstract}
We present a direct comparison between interface-resolved and one-way-coupled point-particle direct numerical simulations (DNS) of gravity-free turbulent channel flow laden with small inertial particles, with high particle-to-fluid density ratio and diameter of about $3$ viscous units. The most dilute flow considered, solid volume fraction $O(10^{-5})$, shows the particle feedback on the flow to be negligible, whereas differences with respect to the unladen case, noteworthy a drag increase of about $10\%$, are found for a volume fraction $O(10^{-4})$. This is attributed to a dense layer of particles at the wall, caused by turbophoresis, flowing with large particle-to-fluid apparent slip velocity. The most dilute case is therefore taken as the benchmark for assessing the validity of a widely used point-particle model, where the particle dynamics results only from inertial and nonlinear drag forces. In the bulk of the channel, the first- and second-order moments of the particle velocity from the point-particle DNS agree well with those from the interface-resolved DNS. Close to the wall, however, most of the statistics show major qualitative differences. We show that this difference originates from the strong shear-induced lift force acting on the particles in the near-wall region. This mechanism is well captured by the lift force model due to \citeauthor{Saffman-JFM-1965} (\emph{J.\ Fluid Mech.}, vol.\ 22 (2), 1965, pp.\ 385--400), while other widely used, more elaborate, approaches aiming at extending the lift model for a wider range of particle Reynolds numbers can actually underpredict the magnitude of the near-wall particle velocity fluctuations for the cases analysed here.
\end{abstract}


\section{Introduction}\label{sec:intro}
Turbulent flows laden with small inertial particles are found in many environmental and industrial contexts. These flows are inherently chaotic and multi-scale, with the interphase coupling categorized by the relevance of the dispersed phase to the overall dynamics \citep{Elghobashi-ASR-1994}. The so-called \emph{one-way}-coupling regime corresponds to low volume and mass fractions of the solid phase, when particle--fluid interactions are negligible, and particle--particle interactions unlikely. Increasing the solid mass fraction while keeping the volume fraction at the same order of magnitude  results in a regime where the overall particle load becomes high enough to modulate the turbulent flow, while particle--particle interactions remain negligible -- the \emph{two-way}-coupling regime. Finally, further increasing the volume fraction results in a regime where particle--particle interactions are also important -- the \emph{four-way}-coupling regime. From a modelling perspective, another important distinction concerns the particle size. When the ratio between the particle size and the Kolmogorov scale is smaller than one, the term \emph{point particle} is used, and particle--fluid coupling is considered to take place at a single point. Conversely, when the size ratio is large, the particles are termed \emph{finite sized} \citep{Balachandar-and-Eaton-ARFM-2010}.\par
Particle-laden turbulence in the one-way-coupling and point-particle limit has been the subject of numerous studies throughout the last decades \cite[for recent reviews see e.g.][]{Toschi-and-Bodenschatz-ARFM-2009,Balachandar-and-Eaton-ARFM-2010}. In these cases, it is assumed that the local  properties of an \emph{undisturbed} flow at the particle position drive the dispersed-phase dynamics \citep{Maxey-Riley-PoF-1983,Gatignol-1983}. For relatively high particle-to-fluid density ratios, the particle dynamics is often simplified to a balance between particle inertial and drag forces, where the latter is a function of the so-called particle-to-fluid slip velocity. Under these conditions,  particles display preferential clustering even in homogeneous and isotropic flows \citep{Toschi-and-Bodenschatz-ARFM-2009}. When the flow is inhomogeneous, particles tend to migrate from regions of high to low turbulence intensity due to turbophoresis \citep{Reeks-JAS-1983}. In turbulent wall-bounded flows, in particular, the particle distribution is driven by the interplay between small-scale clustering, turbophoresis, and the interaction between the particles and near-wall turbulence structures \citep{Soldati-and-Machioli-IJMF-2009,Sardina-et-al-JFM-2012}. When the system reaches a statistical equilibrium, particles tend to accumulate in the low-speed regions near the wall, resulting in a very inhomogeneous local particle concentration; see e.g.\ \cite{Fessler-et-al-PoF-1994,Uijttewaal-and-Oliemans-PoF-1996,Marchioli-et-al-IJMF-2003,Marchioli-et-al-IJMF-2008,Kuerten-PoF-2006}.\par
In flows with locally higher mass loading, two-way-coupling effects may become important. In addition to solving the particle dynamics, a two-way-coupling point-particle algorithm must impose a localized momentum source/sink corresponding to the particle back-reaction to the flow. The challenge here is to determine the local characteristics of the undisturbed flow field, while the flow itself is subjected to a local disturbance due to the presence of the dispersed phase \citep[see e.g.][]{Gualtieri-et-al-JFM-2015}. The classical approach is the particle-in-cell method, developed by \cite{Crowe-et-al-JFE-1977}. Although widely used, the success of this method strongly depends on the number of particles per grid cell, i.e.\ it does not converge with grid refinement. Approaches for a consistent and more robust treatment are the object of active research, as shown by the number of recent studies, e.g.\ \cite{Gualtieri-et-al-JFM-2015,Horwitz-and-Mani-JCP-2016,Ireland-and-Desjardins-JCP-2017}. We should also note the recent efforts to account for pairwise particle--particle hydrodynamic interactions in point-particle models \citep{Akiki-and-Balachandar-JCP-2017}. Investigations of  particle-laden turbulent flows in the two-way-coupling regime are found in e.g.\ \cite{Vreman-et-al-FTC-2009,Capecelatro-and-Desjardins-JFM-2018}.\par
Despite the numerous studies involving direct numerical simulations (DNS) with point-particle methods, validations of the underlying assumptions remain scarce, even for simple flows. For instance, the parameter range where the Maxey-Riley-Gatignol equations are valid remains elusive \citep{Bergougnoux-et-al-PoF-2014}. Though experiments in particle-laden turbulence are insightful \citep{Eaton-and-Fessler-IJMF-1994,Kaftori-and-Banerjee-PoF-1995}, parameter-matched numerical simulations remain challenging, because of the numerical limitations in terms of Reynolds number and the need for experiments in well-controlled, often idealized, configurations. Recent efforts in this direction have been undertaken in the recent study by \cite{Wang-et-al-IJMF-2019}.\par
Another possible reference for new models is a particle-resolved (also denoted interface-resolved) DNS, i.e.\ a DNS that resolves the flow around the surface of each small particle. Despite the great computational challenge of such computations, the first direct comparisons between point-particle models and particle-resolved simulations have started to appear for forced homogeneous isotropic turbulence (HIT) with fixed particles \citep{Vreman-JFM-2016} and decaying HIT with moving particles \citep{Schneiders-et-al-F-2017,Schneiders-et-al-JFM-2017,Mehrabadi-et-al-JFM-2018,Frohlich-et-al-FTC-2018}.\par
We consider particle-laden turbulent channel flow, a widely studied case with minimal governing parameters and particularly important to benchmark models for wall-bounded particle transport. Though interface-resolved DNS of these flows are in general quite demanding, recent studies have demonstrated that massively parallel simulations of wall-bounded flows with $O(10^6)$ interface-resolved particles and $O(10^9-10^{10})$ grid points have become feasible \cite[see][]{Costa-et-al-PRL-2016,Kidanemariam-and-Uhlmann-JFM-2017,Horne-and-Manesh-JCP-2019}.\par
We present interface-resolved DNS of gravity-free turbulent channel flow laden with small inertial particles (with a size of $3$ viscous units, and $100$ times denser than the fluid), in the dilute regime. Two cases are considered, with bulk volume solid fractions that approach the one-way-coupling regime: $0.003\%$ and $0.03\%$. Both cases show a turbophoretic particle drift towards the wall, as expected. However, due to the inhomogeneous particle distribution, non-negligible two-way-coupling effects are found for the case with the largest volume fraction, but not in the most dilute case. The interface-resolved DNS are complemented with corresponding one-way-coupled point-particle DNS at the same Reynolds number using models for the drag and lift forces. Considering only the balance between inertia and nonlinear drag, the turbophoretic wall concentration is overestimated by a factor of $2-3$ with respect to the most diluted case considering fully resolved particles.
We show that the missing ingredient is a shear-induced lift force that reduces the turbophoretic drift. Finally, we demonstrate that these dynamics can be well captured using the model proposed in the seminal work of \cite{Saffman-JFM-1965}.
\section{Methods and computational set-up}\label{sec:methods}
\begin{table}
  \begin{center}
  \def~{\hphantom{0}} \begin{tabular}{cllll} \ct{Case}   & \ct{$\Phi \,\,\,\,(N_p)$} & \ct{$\Psi$} & Notes & $\Rey_\tau$\\[3pt]
      \textit{VD} & $0.003\%\,\,(  500)~~~$   & $0.337\%$   & interface-resolved  (VD)          & $180 \pm 1$\\
      \textit{D~} & $0.034\%\,\,(5\,000)~~$   & $3.367\%$   & interface-resolved  (D)           & $188 \pm 1$\\
      \textit{PP} & \ct{--}                   &  \ct{--}    & point-particle (one-way-coupling) & $179$
  \end{tabular}
      \caption{Computational parameters of the DNS dataset. $\Phi$/$\Psi$ denotes the bulk solid volume/mass fraction, and $N_p$ the total number of particles. For all cases, the bulk Reynolds number $\Rey_b = 5\,600$ (i.e.\ unladen friction Reynolds number $\Rey_\tau^{sph}\approx 180$); particle size ratio $D/(2h)=1/120$; particle-to-fluid mass density ratio $\Pi_\rho=100$, corresponding to a particle diameter (in viscous units)  $D^+=3$ and a Stokes number $St=50$. For the interface-resolved cases, the fluid domain is discretized on a regular Cartesian grid with $(L_x/N_x)\times (L_y/N_y)\times (L_z/N_z) = (6h/4320)\times (2h/1440)\times (3h/2160)$, while the particles are resolved with $D/\Delta x=12$ grid points over the particle diameter ($420$ Lagrangian grid points in total). For case \textit{PP}, $500\,000$ point particles have been simulated in a grid which is $5$ times coarser in each direction. Note that the last column reports the friction Reynolds number $\Rey_\tau$ extracted from the simulations.}
  \label{tbl:comput_params}
  \end{center}
\end{table}
The numerical method solves the continuity and Navier-Stokes equations for an incompressible Newtonian fluid with density $\rho_f$ and dynamic viscosity $\mu$ (kinematic viscosity $\nu=\mu/\rho_f$),
\begin{align}
\boldsymbol{\nabla}\cdot\mathbf{u}       &= 0 \mathrm{,} \label{eqn:cont} \\
    \rho_f\frac{\mathrm{D}\mathbf{u}}{\mathrm{D}t} &= \nabla\cdot\boldsymbol{\sigma}\mathrm{;}\label{eqn:mom}
\end{align}
where $\mathbf{u}$ is the fluid velocity vector, $\boldsymbol{\sigma} = -(p+p_e) \mathbf{I} + \mu\left(\nabla {\mathbf u} + \nabla {\mathbf u}^T \right)$, with $p+p_e$ being the fluid pressure with respect to an arbitrary constant reference value; the resulting term $\boldsymbol{\nabla} p_e$ in eq.~\eqref{eqn:mom} corresponds to a uniform pressure gradient that may serve as driving force for the flow.
\subsection*{Interface-resolved particle simulations}
The fluid equations are coupled to the Newton-Euler equations governing the motion of a spherical particle with mass $m$, and moment of inertia $I$, 
\begin{align}
    m \dot{\mathbf{U}} &= \oint_{\partial V} \! \boldsymbol{\sigma} \cdot \mathbf{n} \, \mathrm{d} A + \mathbf{F}_c \mathrm{,} \label{eqn:nwtn} \\
    I \dot{\boldsymbol{\Omega}} &=\oint_{\partial V} \! \mathbf{r} \times (\boldsymbol{\sigma} \cdot \mathbf{n}) \, \mathrm{d} A + \mathbf{T}_c \mathrm{,} \label{eqn:eulr}
\end{align}
where we use Newton's dot notation for time differentiation. Here $\mathbf{U}$ and $\boldsymbol{\Omega}$ denote the particle linear and angular velocity vectors, $\mathbf{r}=\mathbf{x}-\mathbf{X}$ the position vector with respect to the particle centroid, $\mathbf{n}$ the outward-pointing unit vector normal to the particle surface $\partial V$, $A$ the surface area of the particle, and $\mathbf{F}_c$ and $\mathbf{T}_c$ correspond to external forces and torques associated with short-range inter-particle or particle--wall interactions (such as solid-solid contact).\par
Equations~(\ref{eqn:cont}--\ref{eqn:mom}) and~(\ref{eqn:nwtn}--\ref{eqn:eulr}) are coupled through the imposition of no-slip and no-penetration boundary conditions at the particle surface:
\begin{equation}
\mathbf{U} + \boldsymbol{\Omega}\times\mathbf{r} \stackrel{!}{=} \mathbf{u} \,\,\,\, \forall \,\,\,\, \mathbf{x} \in \partial V\mathrm{.} \label{eqn:bcs}
\end{equation}\par
The Navier-Stokes equations are solved with a second-order finite-difference method on a three-dimensional, staggered Cartesian grid, using a fast-Fourier-transform-based pressure-projection method \citep{Kim-and-Moin-JCP-1985}. The solver was extended with a direct forcing immersed-boundary method (IBM) for particle-laden flows developed by \cite{Breugem-JCP-2012} \cite[see also][]{Uhlmann-JCP-2005} and the lubrication/soft-sphere collision model for short-range particle--particle and particle--wall interactions in \cite{Costa-et-al-PRE-2015}. Several recent studies describe the method, present validations and assess its computational performance~\citep{Picano-et-al-JFM-2015,Costa-CAMWA-2018,Motta-et-al-CaF-2019}.\par
Turbulent channel flow is simulated in a domain periodic in the streamwise ($x$) and spanwise ($z$) directions, with no-slip/no-penetration boundary conditions imposed at the walls ($y=h\mp h$), where $h$ is the channel half-height. The flow is driven by a uniform pressure gradient that ensures a constant bulk velocity. The physical and computational parameters are reported in table~\ref{tbl:comput_params}. Since the IBM requires a fixed, regular Eulerian grid, resolving the spherical particles with $O(10)$ grid points over the diameter is, by far, what dictates the grid resolution. Hence, the spatial resolution for the interface-resolved simulations is, in each direction, about one order of magnitude larger than what is required for single-phase simulation, leading to $\sim10^{10}$ grid points. 
The solid-solid contact between the rigid particles is frictionless, with a normal dry coefficient of restitution, i.e.\ the ratio of rebound to impact velocity in the absence of lubrication effects $e_{n,d}=0.97$. As suggested in \cite{Costa-et-al-PRE-2015}, the particle--particle/wall collision time is set to  $N_c\approx 8$ times the time step of the Navier-Stokes solver. Lubrication corrections for the normal lubrication force when the distance between two solid surfaces is small are used as described in \cite{Costa-et-al-PRE-2015}, with the same model parameters as those used in \cite{Costa-et-al-JFM-2018}.\par
To mimic a flow close to the one-way-coupling regime, we consider two low values of solid volume fraction, $\Phi\simeq3\cdot10^{-5}$, denoted very dilute (\textit{VD}), and $\Phi\simeq 3\cdot10^{-4}$, dilute (\textit{D}). The bulk Reynolds number $\Rey_b = U_b(2h)/\nu=5\,600$, which corresponds to an unladen friction Reynolds number $\Rey_\tau^{sph} = u_\tau h/\nu\approx 180$; where $U_b$ is the bulk flow velocity and $u_\tau$ the wall friction velocity. The particle properties are chosen close to those used in point-particle simulations at the same $\Rey_\tau$, which result in strong turbophoresis and large-scale clustering \citep{Sardina-et-al-JFM-2012}. This choice corresponds to a particle Reynolds number $\Rey_p = D u_\tau/\nu =D^+= 3$, and Stokes number $St_p = \Pi_\rho \Rey_p^{2}/18=50$, where $D$ is the particle diameter and $\Pi_\rho$ the particle-to-fluid mass density ratio.
Two strategies have been considered for achieving an initial condition close to the fully developed turbulent state with little computational cost: (1) using a fully developed one-way-coupled point-particle simulation at a much coarser grid, interpolated onto the grid of the interface-resolved case; and (2) using the fully developed state of a coarse interface-resolved DNS, where the fluid is well resolved, but the particles are under-resolved ($4$ grid points over the particle diameter). The latter approach was preferred, as both the particle distribution and two-way-coupling effects are closer to the fully developed turbulent state of the interface-resolved DNS.\par
Finally, the mesoscale-averaged profiles reported in this manuscript correspond to mean intrinsic averages. The intrinsic average of an observable $o$ is given by
\begin{equation}
 \langle o\rangle(y_j) = \frac{\sum_{ik,t}o_{ijk,t}C_{ijk,t}}{\sum_{ik,t}C_{ijk,t}}\mathrm{,}
\end{equation}
where $C_{ijk,t}$ is the volume fraction of the specific phase at the grid cell $ijk$ and instant $t$, and $y_j$ the wall-normal location of the averaging bin, which extends over the entire domain in the two homogeneous directions. These quantities have been obtained from an ensemble average on about $200$ instances sampled over a period of $100 h/U_b$ once a fully developed state has been reached. The velocity statistics of the solid phase consider the rigid-body motion throughout the volume of the particles, i.e.\ account for both translation and rotation. We should note that higher-order particle statistics pertaining to case \textit{VD} still show some high-frequency fluctuation away from the wall, due to the extremely low local volume fraction.\par 
Each interface-resolved simulation cost about $10$ million core hours on the supercomputer Marconi based in Italy at CINECA.
\subsection*{One-way-coupled point-particle simulations}
The interface-resolved simulations are complemented with one-way-coupled point-particle simulations. Given the large density ratio, we first assume that the particle dynamics simplifies to a balance between inertial and nonlinear Schiller-Naumann \citep{Schiller-Naumann-1933} drag forces, as often done in the literature. Test simulations confirmed that the results are not sensitive to Fax\'en corrections in the particle dynamics. In addition to this standard case, denoted \textit{PP} (point particle), we also investigate the role of shear-induced lift forces, which may be important close to the wall.\par
The point-particle dynamics is driven by the conservation of linear momentum of the particle,
\begin{align}
    m \dot{\mathbf{U}} &= \mathbf{F}_d+\mathbf{F}_l\mathrm{,}\\
      \dot{\mathbf{X}} &= \mathbf{U}\mathrm{,}
\end{align}
where
\begin{align}
    \mathbf{F}_d &= -3\pi\mu D \mathbf{U}_s\left(1+0.15\Rey_p^{0.687}\right)\mathrm{,}\, \mathrm{and}\label{eqn:drag}\\
    \mathbf{F}_l &= 1.615 J\mu D|\mathbf{U}_s|\sqrt{\frac{D^2|\boldsymbol{\omega}|}{\nu}}\frac{\boldsymbol{\omega}\times\mathbf{U}_s}{|\boldsymbol{\omega}||\mathbf{U}_s|}\mathrm{;}\label{eqn:lift}
\end{align}
with $\mathbf{U}_s=\mathbf{U}-\mathbf{u}|_{\mathbf{x}=\mathbf{X}}$ the particle-to-undisturbed-flow slip velocity, and $\boldsymbol{\omega}=\boldsymbol{\omega}|_{\mathbf{x}=\mathbf{X}}$ the undisturbed flow vorticity evaluated at the particle position. Two widely used approaches are considered for modelling the lift force $\mathbf{F}_l$. First, $\mathbf{F}_l$ is modelled simply by the Saffman lift force \citep{Saffman-JFM-1965}, i.e.\ $J=1$ in eq.~\eqref{eqn:lift}. Second, we consider a correction for finite-Reynolds-number effects that fits the tabulated results in \cite{Mclaughlin-JFM-1991} proposed by \cite{Mei-IJMF-1992}, where $J$ in eq.~\eqref{eqn:lift} is a function of a parameter $\varepsilon=\sqrt{|\boldsymbol{\omega}|\nu}/|\mathbf{U}_s|$ \cite[note that the formula in the original reference has a typo; cf.][]{Loth-and-Dorgan-EFM-2009}:
\begin{equation}
    J = 0.3\left(1+\tanh{\left[\frac{5}{2}\left(\log_{10}\varepsilon+0.191\right)\right]}\right)\left(\frac{2}{3}+\tanh\left(6\varepsilon-1.92\right)\right)\mathrm{.}
\end{equation}
We should note that this type of correction for finite-Reynolds-number effects has been employed in numerous studies, and often as basis for more elaborate models \citep[see e.g.][]{Uijttewaal-and-Oliemans-PoF-1996,Wang-et-al-IJMF-1997,Marchioli-and-Soldati-JFM-2002,Marchioli-et-al-IJMF-2007,Loth-and-Dorgan-EFM-2009}. The results obtained using the simple Saffman lift term are here denoted \textit{PP-Saffman}, and those obtained with the correction as \textit{PP-McLaughlin}.\par
In all point-particle simulations, the fluid observables evaluated at the particle position are computed using trilinear interpolation, and the particle positions integrated in time with the same third-order low-storage Runge-Kutta (RK) scheme used for the interface-resolved simulations. As regards particle-wall collisions, a perfectly elastic hard-sphere rebound is adopted: if a particle is about to overlap with the wall during an RK substep with size $\Delta t_s$, the particle final wall-normal position $Y$ and velocity $V$ are computed analytically from the two-body kinematics assuming a piecewise-constant particle velocity. For instance, for a particle colliding onto the lower wall ($y=0$) with centroid position $Y_{in}$ and impact wall-normal velocity $V_{in}<0$, we have, \emph{if} $Y_{in}+V_{in}\Delta t_s <D/2$,
\begin{align}
   V &= -V_{in}\mathrm{,}\\
   Y &= V\Delta t_s - (Y_{in} - D/2)\mathrm{.}
\end{align}
\section{Results}\label{sec:results}
\subsection*{Interface-resolved simulations and the role of lift force}
\begin{figure}
    \centering
    \includegraphics[width=0.99\textwidth]{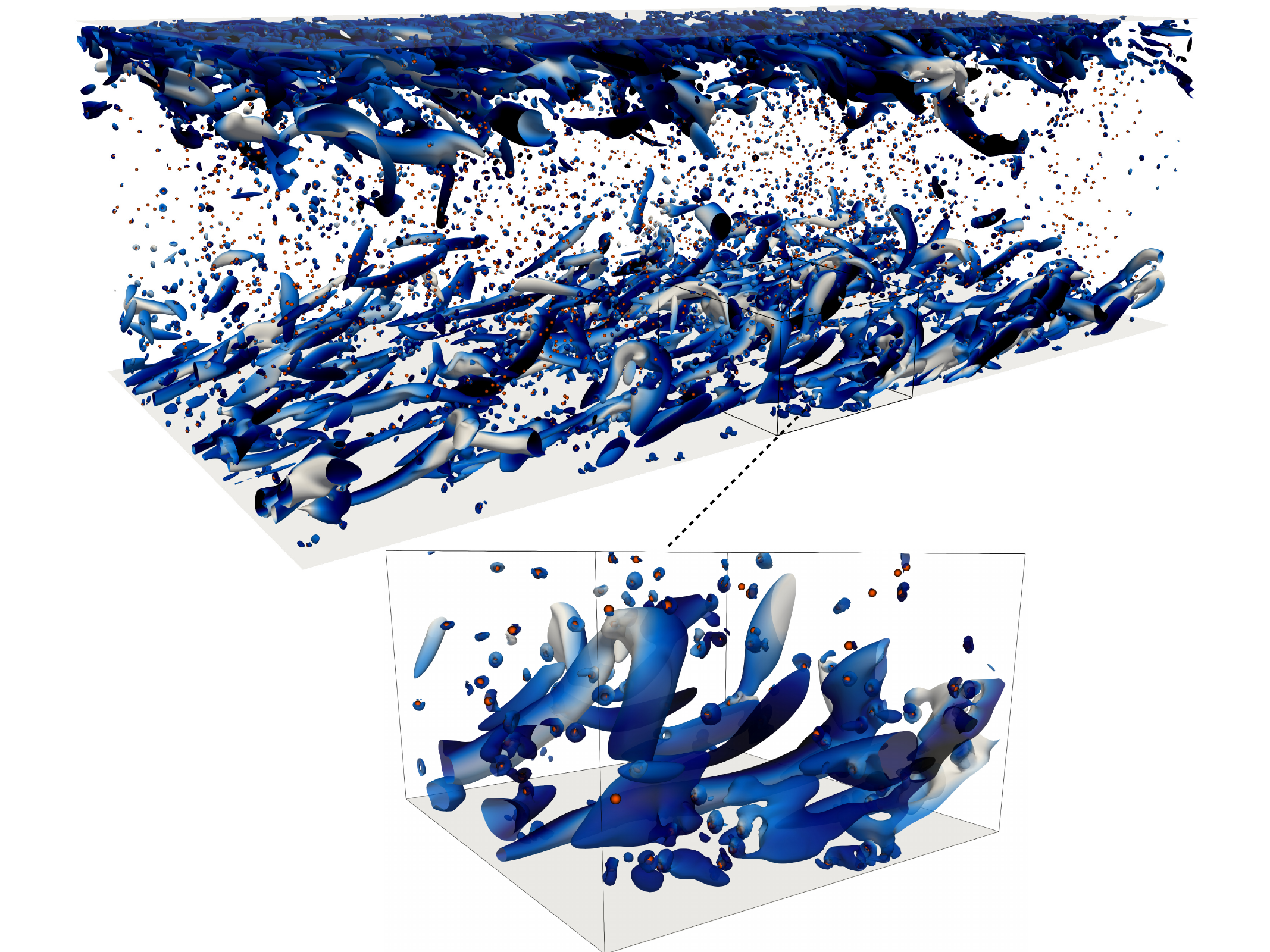}\hfill
    \caption{Visualization of case \textit{D}. Isocontours of the Q-criterion $Q=20(U_b/h)^2$ \citep{Hunt-et-al-1988} coloured by the local wall-normal velocity $v$ (increasing from blue to white with minimum/maximum values of $\mp 0.1U_b$). The interface-resolved particles are depicted in orange. The isosurfaces have been rendered from a field in a grid $4$ times coarser in each direction than that used in the DNS.}
\label{fig:visu_full}
\end{figure}
\begin{figure}
    \centering
    \includegraphics[width=0.99\textwidth]{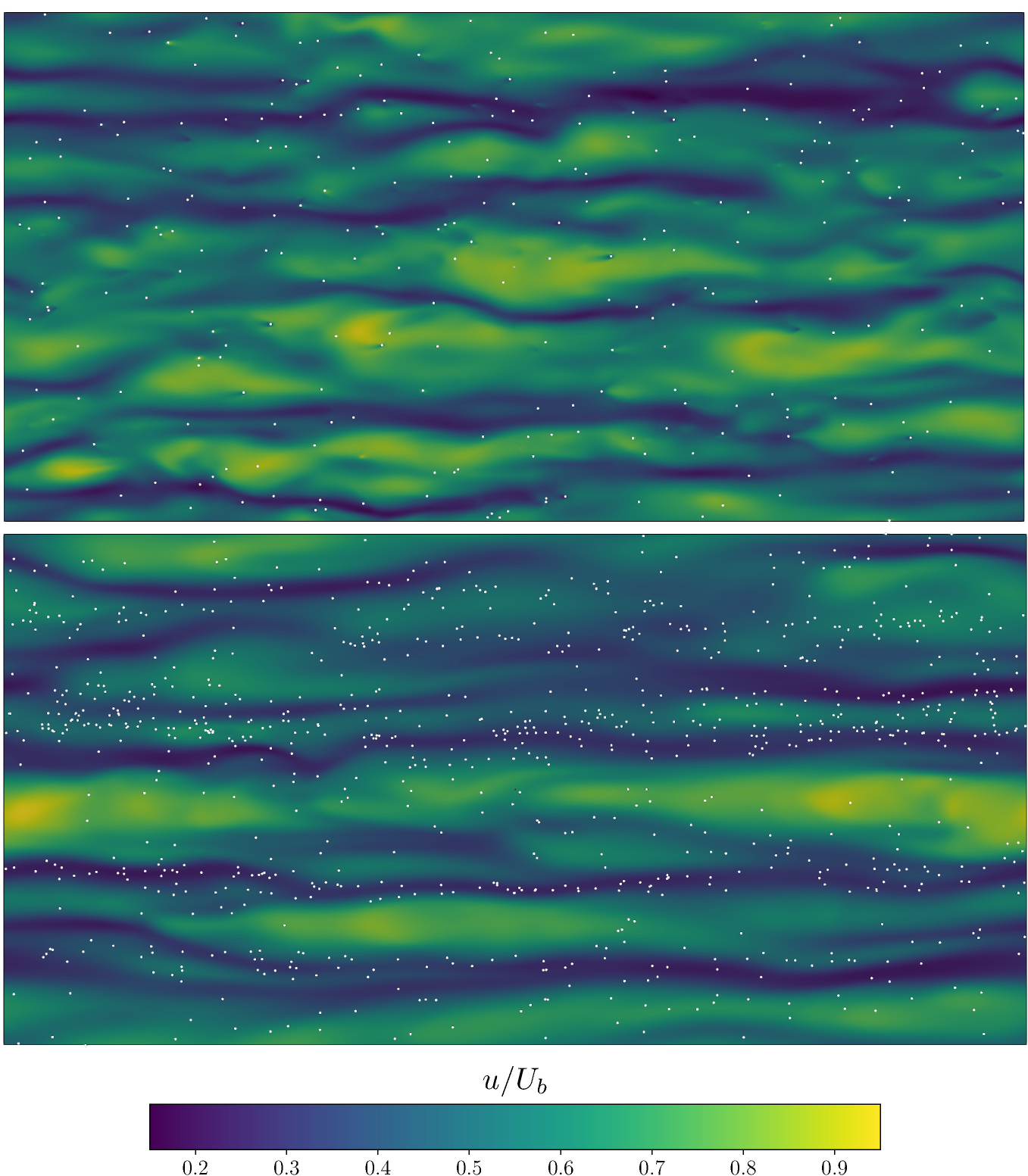}\hfill
    \caption{Contours of streamwise fluid velocity (flow from left to right) in the plane $y/h=0.056$, i.e.\  $y^+ \approx 10$. Particles with wall distance $y_p/h<0.06$ are depicted in white colour. The top and bottom panels correspond to cases \textit{D} and \textit{PP} with matched bulk number density.}
\label{fig:visu_plane}
\end{figure}
\begin{figure}
   \centering
    \includegraphics[width=0.49\textwidth]{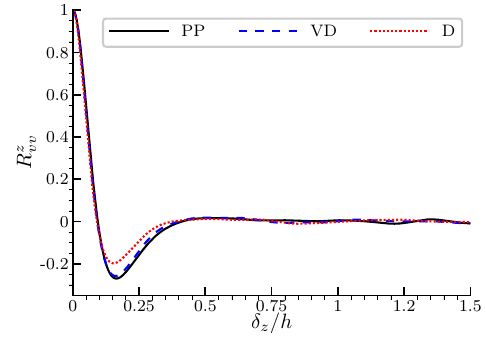}
    \includegraphics[width=0.49\textwidth]{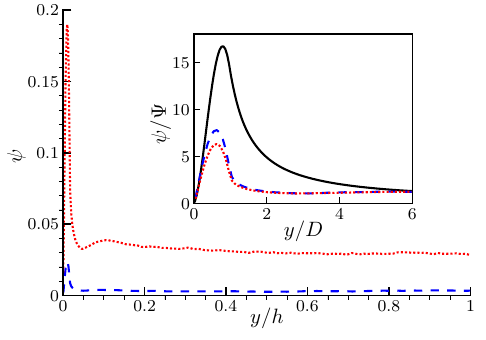}
    \put(-185,120){(\textit{b})}
    \put(-370,120){(\textit{a})}
    \caption{  (\textit{a}) Spanwise autocorrelations of the wall-normal velocity $R_{vv}^z$, at $y/h=0.056$, i.e.\ $y^+\approx 10$.  (\textit{b}) Local solid mass fraction as a function of the outer-scaled wall-normal distance (same legend as the former panel). The inset shows the corresponding bulk-normalized profile, versus the wall-distance in particle diameters.}\label{fig:ruuz_out_phase}
\end{figure}

Figure~\ref{fig:visu_full} shows a visualization of the particle-laden flow for case \textit{D}. Similarly to case \textit{VD} (not shown), the near-wall large-scale structures resemble those of the unladen flow. Nonetheless, the localized effect of the particles is evident, as depicted by the high-vorticity trail due to their wakes. The overall drag is a measure of the cumulative effect of the localized disturbances: this is reported in the last column of table~\ref{tbl:comput_params} in terms of a friction Reynolds number $\Rey_\tau = u_\tau h/\nu$. 
While the very dilute (\textit{VD}) case shows, as expected, about the same drag as the unladen flow, the dilute case (\textit{D}) shows about $5\%$ higher friction Reynolds number (i.e.\ about $10\%$ increase in pressure drop). This is a remarkable drag increase, better comparable to what has been observed for finite-sized neutrally buoyant particle suspensions with volume fraction of $5\%$ \citep{Picano-et-al-JFM-2015}. The modulation of the near-wall structures for case \textit{D} is noticeable in figure~\ref{fig:visu_plane}, where we display the streamwise velocity contours close to the wall. In addition to showing a microscopic footprint, the resolved particles disrupt streamwise-correlated near-wall structures, as can be noticed in the spanwise autocorrelation of the wall-normal fluid velocity, given by $R_{vv}^z\left(\delta_z\right) = \left<v^\prime(z)v^\prime(z+\delta_z)\right>/\left<v^{\prime 2}(z)\right>$ shown in figure~\ref{fig:ruuz_out_phase}(\textit{a}). Clearly, the magnitude of the negative peak in correlation -- footprint of the near-wall low- and high-speed streaks -- is dampened for case \textit{D}, while remaining very close to that of single-phase flow in case \textit{VD}. We will see that these observations are closely connected to the inhomogeneous particle distribution near the wall.\par 
Figure~\ref{fig:ruuz_out_phase}(\textit{b}) depicts the wall-normal profiles of the local solid mass fraction as a function of the wall-normal distance. The inset shows the same quantity divided by the bulk values versus the wall-normal distance in particle diameters. The profiles show a near-wall peak at $y\approx D/2$ about one order of magnitude larger than the bulk value, which explains the drag increase observed in case \textit{D}. Despite the low bulk mass loading, the near-wall mass fraction becomes high enough to modulate the flow in this critical region ($\Psi \simeq 0.2$). We should note that the corresponding local volume fraction is still too low ($\Phi \simeq 0.002$) for particle--particle interactions to be significant: 
the collision frequency in the viscous sublayer is estimated to be less than $10^{-5}$ collisions per particle pair per bulk time unit $h/U_b$ for the less dilute case \textit{D}. The interface-resolved simulations show pronounced differences when compared to the point-particle results (\textit{PP}), which consider only the nonlinear drag in the dynamics. In particular, the latter overpredicts the concentration peak by a factor of $2-3$. Moreover, the concentration profile of case \textit{PP} shows
a more gentle decrease away from the wall, while in the other cases the concentration peak corresponds to a single particle layer. We attribute this localized peak to the combination of turbophoresis, the kinematic constraint that the wall imposes on the particles, and the stabilizing effect of lubrication forces \citep{Picano-et-al-JFM-2015}. Clearly the mechanisms causing preferential concentration in the standard point-particle simulation are greatly weakened. At least for the very dilute case \textit{VD}, these differences cannot be explained in terms of turbulence modulation nor particle--particle interactions (i.e.\ two/four-way-coupling effects). Hence, the cause must be the different near-wall dynamics of (isolated) particles.\par
\begin{figure}
   \centering
   \includegraphics[width=0.49\textwidth]{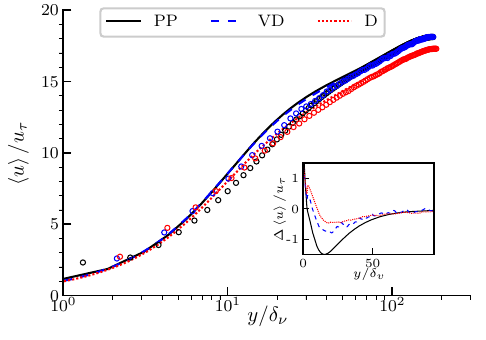}\hfill
   \includegraphics[width=0.49\textwidth]{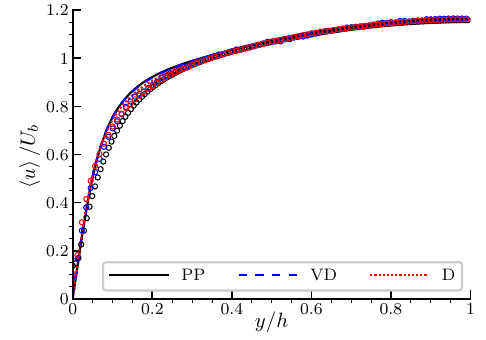}
    \put(-183,120){(\textit{b})}
    \put(-380,120){(\textit{a})}
    \caption{(\textit{a})  Inner- and (\textit{b}) outer-scaled mean velocity profiles for the different cases (lines -- fluid velocity; symbols (colour-matched) -- particle velocity). The inset in panel (\textit{a}) shows the inner-scaled difference between fluid and particle velocity profiles $\Delta\left<u\right>$.}\label{fig:umean}
\end{figure}

Figure~\ref{fig:umean}(\textit{a}) shows the inner-scaled profiles of mean streamwise fluid and particle velocity for the different cases. While the fluid velocity profile of case \textit{VD} matches that of the unladen flow, case \textit{D} shows significant deviations. Since the inner-scaled profiles of all cases agree in the outer region (see panel (\textit{b})), these deviations are attributed to the increase in wall shear. We thus confirm that, somewhat unexpectedly, only case \textit{VD} satisfies the one-way-coupling assumption, i.e.\ the turbulence modulation by the dispersed phase is negligible. This case serves therefore as benchmark to assess the validity of the point-particle models to predict the dispersed-phase dynamics in the one-way-coupling regime.\par
As for the particles, they tend to flow slower than the fluid in the buffer layer ($10 \lesssim y/\delta_\nu\lesssim 40-50$, with $\delta_\nu = \nu/u_\tau$; see also the figure inset, showing the inner-scaled difference between the profiles of each phase). This velocity difference is more pronounced in case \textit{PP}, where the particle velocity reduction is clear also very close to the wall. This has been observed in previous studies using point-particle DNS~\citep{Sardina-et-al-JFM-2012}, and is attributed to the preferential sampling of the fluid low-speed streaks near the wall. The particle-resolved cases show a much weaker reduction of the mean particle velocity, which suggests that particles reside in the low-speed regions for shorter times before resuspending into the bulk; see also the discussion below about the  particle dynamics. In the viscous sublayer, particles show a slightly higher mean velocity in the interface-resolved simulations. This higher slip velocity causes \textit{hot-spots} of higher wall shear stress, which favour an increase in overall drag \citep{Costa-et-al-PRL-2016,Costa-et-al-JFM-2018}. Clearly this effect is significant in case \textit{D}, where the near-wall number density is high enough, but not in case \textit{VD}.\par
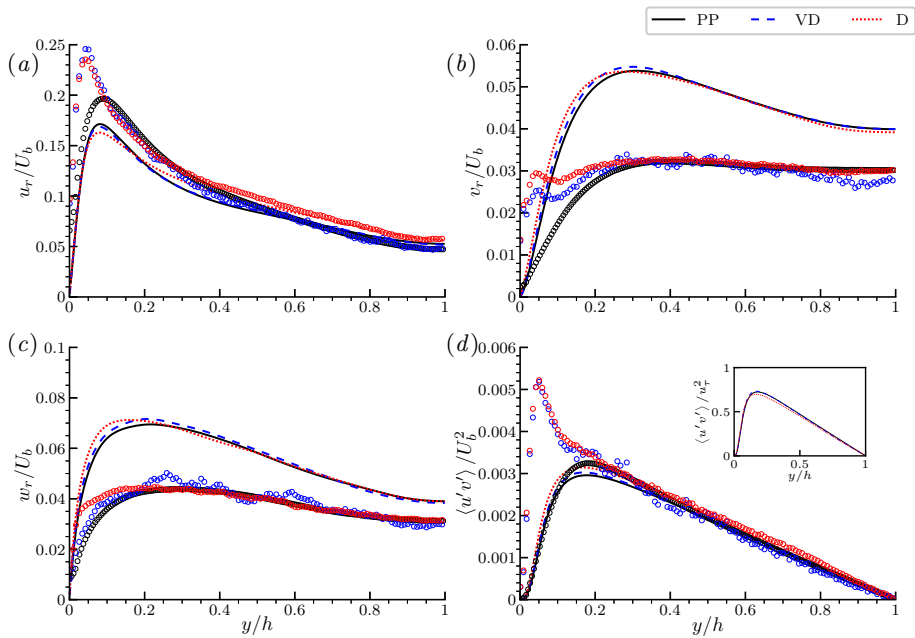
\begin{figure}
   \centering
   \includegraphics[width=0.9\textwidth]{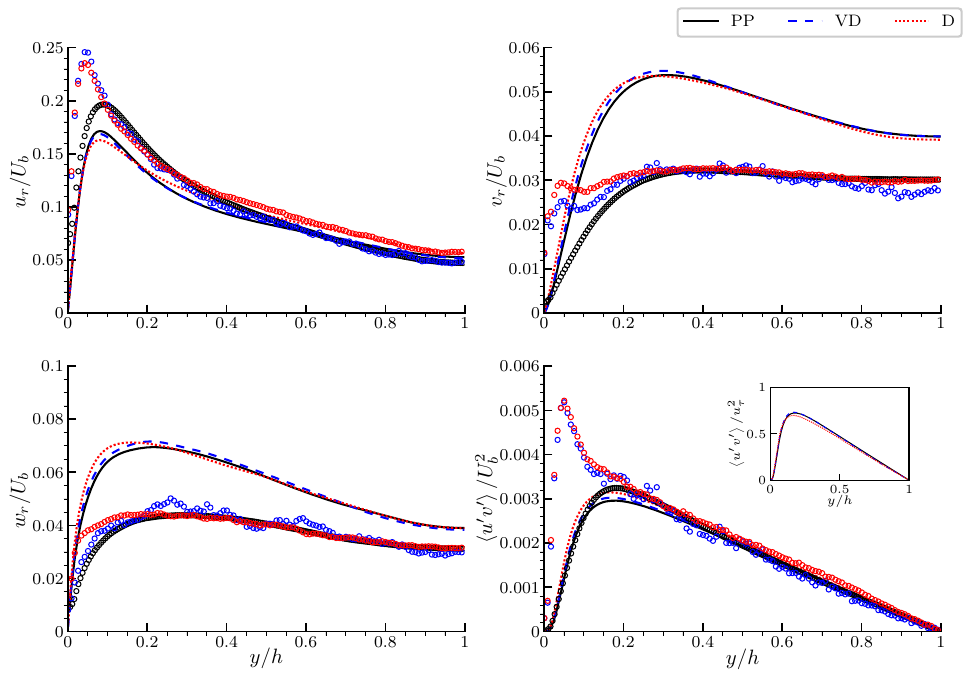}
    \put(-180,110){(\textit{d})}
    \put(-345,110){(\textit{c})}
    \put(-180,215){(\textit{b})}
    \put(-345,215){(\textit{a})}
    \caption{Outer-scaled second-order moments of particle velocity. (\textit{a}) streamwise velocity r.m.s.\; (\textit{b}) wall-normal velocity r.m.s.\;  (\textit{c}) ditto for spanwise velocity. (\textit{d}) Reynolds stresses profile with inner-scaled inset. Lines -- fluid; symbols (colour-matched) -- particles.}\label{fig:velstats}
\end{figure}

Figure~\ref{fig:velstats} shows the second-order statistics of fluid and particle velocity. Focusing first on the fluid phase, we see once more that the data for \textit{VD} tend to those of the single-phase flow, with the minor differences attributed to a better statistical convergence of case \textit{PP}, and possibly a slight two-way-coupling effect. Conversely, turbulence modulation is evident for case \textit{D}. Here the Reynolds stresses are higher, consistently with the overall drag increase (see the figure inset). Moreover, small differences are found for all the velocity root mean square (r.m.s.) values of case \textit{D} near the wall, where the velocity fluctuations become less anisotropic, i.e.\ $u_r$ decreases and $v_r$ and $w_r$ increase. This is attributed to the enhanced mixing due to the near-wall particles, whose local mass fraction is high enough for two-way-coupling effects to be significant. \par
\par
\begin{figure}
   \centering
   \includegraphics[width=0.59\textwidth]{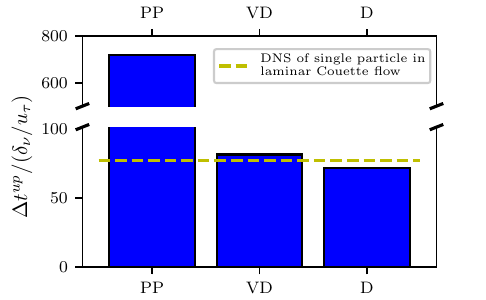}
    \caption{Inner-scaled average time a wall-skimming particle takes to reach a wall-normal distance $y>5\delta_\nu$ (i.e.\ to exit the viscous sublayer), $\Delta t^{up}$. The dashed yellow line correspond to the time a wall-skimming particle takes, in a DNS of a model laminar Couette flow at equivalent Reynolds number, to reach the same inner-scaled wall-normal distance.}\label{fig:res_time}
\end{figure}

Interestingly, the second-order moments of the particle velocity for the fully resolved \emph{one-way}-coupling case, \textit{VD}, strongly differ from those of the point-particle simulations near the wall, while in the bulk the two cases display a similar behaviour. In the bulk, where the local shear is relatively low, the point-particle model succeeds in predicting the particle dynamics. We should note that the same closure for the point-particle dynamics was used in
\cite{Mehrabadi-et-al-JFM-2018} for decaying HIT, and the results also compared well to the corresponding interface-resolved case. Closer to the wall ($y/h \lesssim 0.1$), however, the interface-resolved simulations show higher fluctuation levels than the point-particle reference. One can depict a clear change in trend in case \textit{VD} for the profiles pertaining the quantities in the plane at $y/h\approx 0.1$. There is a clear local minimum of $v_r$, and a sudden change in slope for
$\left<u'v'\right>$. The exception is the spanwise velocity r.m.s.\ $w_r$, which attains similar values also close to the wall for cases \textit{VD} and \textit{PP}. We should note that similar trends for the streamwise and wall-normal particle velocity r.m.s.\ have been observed in recent experiments of particle-laden turbulent downward flow in a vertical channel; see \cite{Fong-et-al-JFM-2019}. All these observations suggest differences in the single-particle dynamics, in particular in the way
particles approach and depart from the wall in the two models. In the one-way-coupled point-particle DNS, the particle dynamics is modelled by a simple drag law without considering shear-induced lift forces. In this case, particles are driven towards the wall with high velocity by turbophoresis. Their inertia prevents resuspension, resulting in long periods of wall accumulation in low-speed regions, while only few of them drift back into the bulk~\citep{Soldati-and-Machioli-IJMF-2009,Sardina-et-al-JFM-2012}. 
Since at equilibrium the net wall-normal particle flux is zero, a large number of particles accumulate at the wall. Conversely, when the flow around particles is resolved, these tend to reside for much shorter times near the wall before re-suspending. Hence, point particles tend to skim along the wall in low-speed streaks for long periods~\citep{Soldati-and-Machioli-IJMF-2009,Sardina-et-al-JFM-2012}, whereas resolved particles show shorter residence times at the wall, quickly take off, and are not preferentially localized in low-speed streaks, see figure~\ref{fig:visu_plane}(\textit{b}). 
This faster cycle explains the larger value of $v_r$ near the wall, and consequently the larger values of $u_r$ and $\left<u'v'\right>$ since the fluctuations are correlated through the mean shear. To better quantify this effect, figure~\ref{fig:res_time} shows the average time that a particle close to the wall (i.e.\ located at $y\approx D/2$) needs to exit the viscous sublayer (i.e.\ to reach a wall-normal position $y>5\delta_\nu$), $\Delta t^{up}$. The figure shows that near-wall particles in the fully resolved cases take about the same time to exit the viscous sublayer, which is about one order of magnitude shorter than that of the point-particle DNS \textit{PP}.\par
The particle dynamics just described suggests that a shear-induced lift force is the missing key ingredient absent in the point-particle model. Such a force plays a very important role in the particle dynamics near the wall, where the mean shear is high \citep{Wang-et-al-IJMF-1997,Soldati-and-Machioli-IJMF-2009}. It is known that a particle flowing near a wall in a shear flow experiences a strong lift force \citep{Saffman-JFM-1965,Mclaughlin-JFM-1991,Cherukat-and-McLaughlin-JCM-1994,Bagchi-and-Balachandar-PoF-2002,Magnaudet-JFM-2003}. A similar mechanism should work for small particles in the viscous sublayer of a turbulent flow.\par
To confirm the nature of the mechanism for particle detachment, we performed an auxiliary DNS of laminar Couette flow at the same particle Reynolds number. The computational domain has size $L_x\times L_y\times L_z=20D\times 10D \times 10D$ with a regular grid where $D/\Delta x = 16$. The boundary conditions are the same as for the turbulent channel flow, except that the flow is now driven by a non-zero streamwise velocity $U_w$ at $y=L_y$. The Reynolds number based on the local shear rate $\dot{\gamma} = U_w/L_y$ and particle size is set to match that of the particle in the viscous sublayer, i.e.\ $\dot{\gamma}D^2/\nu = (u_\tau/\delta_v) D^2/\nu$. A single particle with the same physical properties is placed at the bottom wall, with initial linear and angular velocity conforming to the local flow velocity and vorticity. The flight time  $\Delta t^{up}$ for the particle to detach from the wall and travel $5$ viscous units in $y$ is reported by the dashed line of figure~\ref{fig:res_time}. The measured time is remarkably close to the average value measured in the interface-resolved DNS for the two turbulent cases under consideration. This strongly suggests that the mechanism for particle detachment from the wall is, to first approximation, purely shear-driven.
Moreover, the following scaling considerations suggest that shear-induced lift forces can be important close to the wall. Let us neglect short-range particle--wall interactions and consider the particle dynamics in the viscous sublayer to be modelled by an unbounded linear shear flow. In the limit of vanishing particle Reynolds number, the ratio between the drag and shear-induced lift force scales with $D^+$ (cf.\ eq.~\eqref{eqn:drag} and~\eqref{eqn:lift} with $J=1$ for Saffman lift, and note that in this case $|\boldsymbol{\omega} \times\mathbf{U}_s|=|\boldsymbol{\omega}||\mathbf{U}_s|=(u_\tau^2/\nu)|\mathbf{U}_s|$):
\begin{equation}
    \left|\frac{\mathbf{F}_l}{\mathbf{F}_d}\right|_{{y^+<5}} = \frac{1.615}{3\pi}\sqrt{\frac{D^2|\boldsymbol{\omega}|}{\nu}} = 0.171 D^+\mathrm{,}
\end{equation}
with a proportionality coefficient $O(1)$ in the present set-up. This suggests that shear-induced lift forces cannot be neglected near the wall when $D^+\gtrsim 1$, as these can be as high as the streamwise drag forces. It should be noted, however, that the lift force might remain important for lower values of $D^+$, since particles in the viscous sublayer would tend to flow for a longer time while subjected to this wall-normal force. Away from the wall, instead, the order of magnitude of the lift force should become negligible with respect to that of the viscous drag, just like the other terms in the Maxey-Riley-Gatignol equations that are typically neglected for small inertial particles in wall-bounded turbulent flows \citep[see e.g.][]{Wang-et-al-IJMF-1997}.\par
We have seen that the dynamics of resolved and point particles is similar in the bulk, and thus the drift towards the wall is well described by point-particle methods. When moving close to the wall, the high shear induces strong lift forces that quickly dislodge particles. The lower value of the near-wall volume fraction peak in figure~\ref{fig:ruuz_out_phase}(b) for case \textit{D}, and its shorter average take-off time in figure~\ref{fig:res_time} are consistent with this picture, as the mean wall shear is higher in this case. This wall detachment mechanism is absent in models considering only the  drag, as in the \textit{PP} case, which instead display slow resuspension and particles spending a long time in the near-wall low-speed streaks. In the next section we assess the validity of simple shear-induced lift models to predict the particle dynamics.
\subsection*{Assessment of lift models for point-particle simulations}
The most well-known models proposed to describe a shear-induced lift force in the one-way-coupling point-particle regime introduced at the end of section~\ref{sec:methods} are here compared with fully resolved simulations. In all cases, the standard Eulerian particle statistics for the point-particle models with lift force are compared to those of the model without lift, and the interface-resolved case \textit{VD}.\par
\begin{figure}
   \centering
   \includegraphics[width=0.49\textwidth]{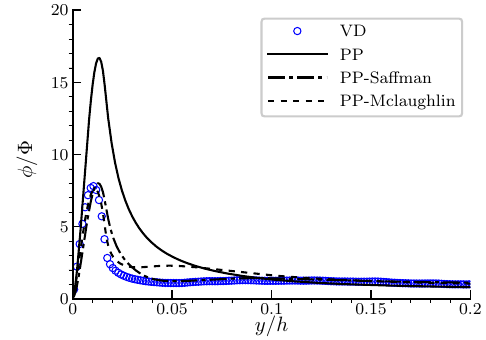}\hfill
   \includegraphics[width=0.49\textwidth]{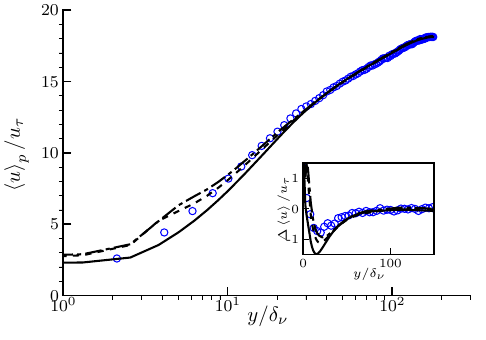}
    \put(-185,120){(\textit{b})}
    \put(-378,120){(\textit{a})}
    \caption{(\textit{a}) Local solid mass fraction as a function of the outer-scaled wall-normal distance for the different point-particle cases. (\textit{b}) Corresponding inner-scaled mean particle velocity profiles. The inset in panel (\textit{b}) shows the inner-scaled difference between fluid and particle velocity profiles $\Delta\left<u\right>$.}\label{fig:phase_umean_parts}
\end{figure}

Figure~\ref{fig:phase_umean_parts}(\textit{a}) shows the normalized local volume fraction profiles for the different cases considered. The point-particle models accounting for lift forces predict well the wall concentration peak. The peak location is predicted slightly away from the wall, probably due to a minor stabilizing effect of short-range particle--wall interactions and softer contact, absent in the point-particle model. These secondary mechanisms can possibly be modelled with near-wall closures such as an effective (\emph{wet}) restitution coefficient that is a function of the particle impact Stokes number \citep[see e.g.][]{Joseph-et-al-JFM-2001,Legendre-et-al-CES-2006,Fong-et-al-JFM-2019}, or by more sophisticated near-wall corrections for the particle dynamics \citep{Gondret-et-al-PoF-2002,Lee-and-Balachandar-JFM-2010}. We should note that the former option may be too simple for realistically predicting the particle dynamics, as the drag force already accounts for part of the fluid effects \citep{Gondret-et-al-PoF-2002}. Indeed, we tested a point-particle simulation with the same dynamics as \textit{PP-Saffman} but with a lower coefficient of restitution of $0.9$; the results showed a shift towards the wall of the concentration peak that, however, overpredicts the maximum concentration by a factor of $1.3$.\par
Away from the wall, the case using the seminal lift model, \textit{PP-Saffman}, optimally predicts the concentration profile. The mean particle velocity pertaining to the cases where a lift force is accounted for is much closer to the interface-resolved case (see figure~\ref{fig:phase_umean_parts}(\textit{b})). This is particularly evident in the apparent mean particle-to-fluid velocity difference displayed in the figure inset, where the results with lift force show a much better agreement with case \textit{VD}, in particular in the region where the negative slip is highest. Overall, both lift models predict relatively well the first-order statistics pertaining to case \textit{VD} shown in figure~\ref{fig:phase_umean_parts}. This is not the case for the second-order moments of the particle velocity.\par
\begin{figure}
   \centering
   \includegraphics[width=0.49\textwidth]{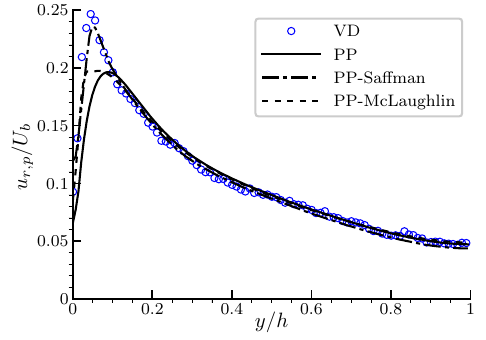}\hfill
   \includegraphics[width=0.49\textwidth]{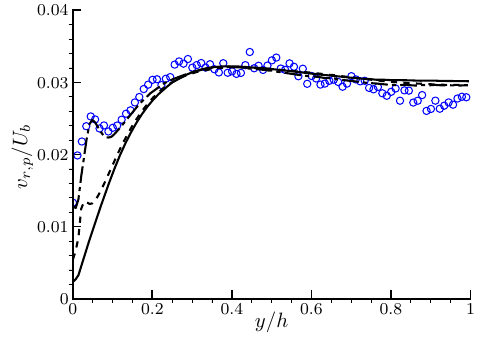}
    \put(-190,122){(\textit{b})}
    \put(-385,122){(\textit{a})}\\
   \includegraphics[width=0.49\textwidth]{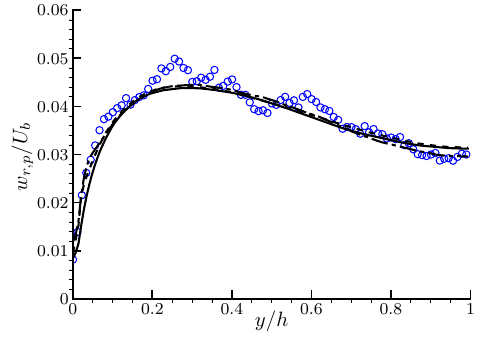}\hfill
   \includegraphics[width=0.49\textwidth]{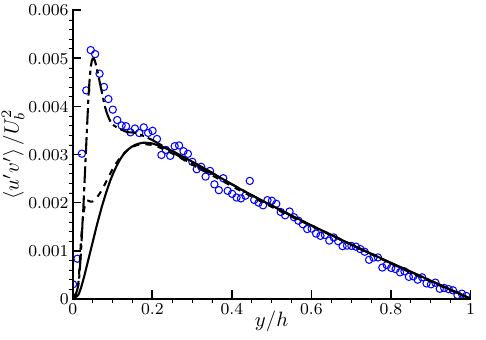}
    \put(-190,122){(\textit{d})}
    \put(-385,122){(\textit{c})}\\
    \caption{Outer-scaled second-order moments of particle velocity for the different point-particle cases. (\textit{a}) streamwise velocity r.m.s.; (\textit{b}) wall-normal velocity r.m.s.; (\textit{c}) ditto for spanwise velocity; (\textit{d}) Reynolds stresses profile.}\label{fig:velstats_parts}
\end{figure}

Figure~\ref{fig:velstats_parts} shows the second-order moments of particle velocity for the different cases considered. As seen in the previous section, the profiles pertaining to the streamwise and wall-normal velocity fluctuations (panels (\textit{a},\textit{b}) and (\textit{d})) of the interface-resolved simulations, case \textit{VD}, show significant differences with case \textit{PP} close to the wall, with a peculiar non-monotonic trend as compared to the smooth decrease obtained from the point-particle simulations. All panels point to a somewhat surprising result: the low-Reynolds-number shear-induced lift model of \cite{Saffman-JFM-1965} predicts almost perfectly the second-order moments of particle velocity, while the model accounting for finite-Reynolds-number effects shows major qualitative differences. Actually, the somewhat better agreement of the Saffman model compared to that of \cite{Mclaughlin-JFM-1991} for interface-resolved simulations of a particle in a shear flow has also been noticed in the work of \cite{Mei-IJMF-1992}. Still, we should stress here the better performance of Saffman's model in this more demanding benchmark, in view of the number of references that have implemented/extended lift models based on the work of \cite{Mclaughlin-JFM-1991} for point-particle simulations. We also tested other approaches that have been suggested in the literature based on the work of \cite{Mclaughlin-JFM-1991} and observed similar discrepancies (not shown here).\par
The very good prediction of Saffman's lift model also sheds light on  the cause of the non-monotonic trend of the fluctuation intensities and shear stresses in panels (\textit{a},\textit{b}) and (\textit{d}) of figure~\ref{fig:velstats_parts} for $y/h<0.1$. In this region, the magnitude of the shear-induced lift force is highest. Hence, a particle approaching the wall is forced to suddenly depart from it, enhancing the velocity fluctuations in the wall-normal and streamwise directions. This mechanism, however, does not increase the spanwise velocity fluctuation.
\par
\section{Conclusions}\label{sec:conclusions}
We have presented two particle-resolved DNS of turbulent channel flow laden with small inertial particles with volume fractions of $3\cdot 10^{-4}$ and $ 3\cdot 10^{-5}$. Since the latter case hardly shows flow modulation, it can be considered to fall into the one-way-coupling assumption, i.e.\ the particles do not influence the statistics of the turbulence. The less dilute case, instead, shows about $10\%$ drag increase compared to the unladen case. This striking increase is attributed to a significant particle mass fraction near the wall caused by turbophoresis. These particles flow with high particle-to-fluid (apparent) slip velocity, producing \textit{hot-spots} of large wall shear.\par
We also examine the differences between the most dilute case, which satisfies the one-way-coupling assumption, and a classic one-way-coupled point-particle simulation where the particle acceleration is balanced by the nonlinear (Schiller-Naumann) drag. While, in the bulk of the channel, concentration profiles and moments of the particle velocity agree well, most of these quantities show clear differences close to the wall. This disagreement is attributed to a missing key ingredient in this point-particle simulation: a shear-induced lift force for the near-wall particle dynamics.
All differences between resolved and point-particle statistics can be related to this near-wall effect. First, the high concentration at the wall in the interface-resolved simulations is limited to a single particle layer, since particles tend to promptly depart from the wall. Second, particles are resuspended before they can accumulate for a long time in low-speed streaks, which lowers the apparent slip velocity in the near-wall region. Finally, because of this lift force, particles show higher near-wall velocity fluctuations.\par
To test this hypothesis, the average particle residence time in the viscous sublayer has been measured and compared to that of a single resolved sphere in a laminar Couette flow with matched (particle shear) Reynolds number. The average flight time of the resolved particles in a turbulent channel is extremely close to that of the laminar simulation, strongly supporting our conclusion. Conversely, point particles reside in the viscous sublayer for a time about one order of magnitude longer. Properly accounting for this reduced residence time is therefore fundamental to accurately predict the particle statistics without resolving the flow conforming to the particles. 
Moreover, a scaling analysis shows that the lift force in the viscous sublayer can be as high as the streamwise drag force: in the limit of vanishing particle Reynolds number, the ratio between the two forces in the viscous sublayer scales with the inner-scaled particle diameter, $D^+$, which is ${O}(1)$ in the cases addressed here.\par
Accordingly, we tested the validity of simple but widely used shear-induced lift models: that from the seminal work of \cite{Saffman-JFM-1965}, and a model based on the work of \cite{Mclaughlin-JFM-1991} that considers finite Reynolds number effects \citep[see][]{Mei-IJMF-1992}. Our results show that both models predict reasonably well the wall peak in particle concentration and mean particle velocity, where the minor differences can be attributed to short-range particle--wall interactions that are not incorporated in the model. As regards the second-order moments of the particle velocity, the interface-resolved results from the simulations show a more complex, non-monotonic trend near the wall for the streamwise and wall-normal r.m.s.\ and the Reynolds shear stress in a region where the point-particle simulation without lift predicts a monotonic decrease. Quite remarkably, the oldest lift model available in the literature, that due to Saffman, quantitatively predicts all these trends, while the other model shows differences with respect to the most dilute interface-resolved simulation. We should remark that we have tested other approaches that have been suggested in the literature based on the work of \cite{Mclaughlin-JFM-1991}, and observed similar discrepancies. The present work opens new perspectives towards a revision of the point-particle models. In particular,  trajectories and force time histories of the point particles could be compared to those of fully resolved particles in order to assess possible local discrepancies and potential improvements. Nevertheless, the  good agreement between statistics of interface-resolved particle simulations and of the point-particle model including the Saffman lift is an indirect proof that the dominant dynamics is well captured. The success of the Saffman lift force model also suggests that short-range particle--wall interactions play a minor role in the near-wall particle dynamics when compared to that of the local shear near the wall. \par
We hope that the present results can be exploited for the development of improved point-particle models for one- and two-way-coupling regimes in wall-bounded turbulent flows.
\section*{Acknowledgements}
We acknowledge {PRACE} for awarding us access to the supercomputer Marconi, based in Italy at {CINECA} under project 2017174185--\textit{DILPART}, and the computing time provided by SNIC (Swedish National Infrastructure for Computing). This work was supported by the European Research Council grant no.\ ERC-2013-CoG-616186, TRITOS, the Swedish Research Council grant no.\ VR 2014-5001, and the grant BIRD192032/19 from the University of Padova. The authors acknowledge the anonymous referees for useful comments on an earlier version of the manuscript. Bert Vreman, Hans Kuerten and Pedram Pakseresht are also thanked for their feedback on an earlier version of the manuscript.
\section*{Declaration of Interests}
None.
\bibliographystyle{jfm}
\bibliography{bibfile.bib}
\end{document}